\documentclass[apjl]{emulateapj}

\usepackage{hyperref}
\usepackage{amsmath}
\usepackage{amssymb}
\usepackage{graphicx}
\usepackage{color}
\usepackage{natbib}
\usepackage[normalem]{ulem}
\usepackage{multibib}
\usepackage{lineno}

\def\be{\begin{equation}}
\def\ee{\end{equation}}
\def\ba{\begin{eqnarray}}
\def\ea{\end{eqnarray}}

\begin{document}

\title{The impact of spin-kick alignment on the inferred velocity distribution of isolated pulsars}
\author{Ilya Mandel\altaffilmark{1,2}}
\email{ilya.mandel@monash.edu}
\author{Andrei P.~Igoshev\altaffilmark{3}}
\affil{$^1$School of Physics and Astronomy, Monash University, Clayton, Victoria 3800, Australia}
\affil{$^2$The ARC Center of Excellence for Gravitational Wave Discovery -- OzGrav, Australia}
\affil{$^3$ Department of Applied Mathematics, University of Leeds, LS2 9JT Leeds, UK}

\begin{abstract}
The speeds of young isolated pulsars are generally inferred from their observed 2-d velocities on the plane of the sky under the assumption that the unobserved radial velocity is not special, i.e., that the measured 2-d velocity is an isotropic projection of the full 3-d velocity.  However, if pulsar spins are preferentially aligned with kicks, then the observer's viewing angle relative to the pulsar velocity vector is in fact special because the direction of the spin impacts the detectability of the pulsar.  This means that the measured 2-d velocity of observable pulsars is not an isotropic projection, which affects inference on 3-d velocities.  We estimate this effect and conclude that it could lead to a $\sim 15\%$ systematic over-estimate of neutron star natal kicks if young pulsars have high obliquity angles and narrow beams, but the exact correction factor depends on the distribution of beam-spin and spin-kick misalignment angles and beam widths.
\end{abstract}

\maketitle

\section{Introduction}

The natal kick velocities of neutron stars play a key role in a number of physical processes, from their retention in globular clusters to the evolution of binaries, ranging from neutron star X-ray binaries to gravitational-wave sources \citep[e.g.,][]{Sigurdsson:2003,BrayEldridge:2016,Igoshev:2021,Mandel:2020}.  Over the past twenty years, a number of studies inferred the pulsar velocity distribution from the increasingly large and accurately measured sample of observed single pulsar velocities \citep[e.g.,][]{Arzoumanian:2002, Hobbs:2005, FGKaspi:2006, IgoshevVerbunt:2017, Igoshev:2020}.  

Generally only the radio pulsar location and the 2-dimensional velocity projected onto the plane of the sky can be measured.   Therefore, most of these studies had to assume that the measured 2-d velocities are isotropic projections of the true 3-d velocity. \citet{IgoshevVerbunt:2017} and \citet{Igoshev:2020} explored semi-isotropic velocity distributions as well: they assumed that the total three-dimensional velocity points away from the Galactic plane for young radio pulsars.    (Because the dynamical timescale in the Galactic potential is $\sim 100$ Myr at our distance of 8 kpc from the Galactic center, acceleration can be neglected for young pulsars with ages $\lesssim 10$ Myr; we therefore also focus on young pulsars in this work.)  Radio pulsar progenitors are massive stars born in the thin Galactic disk, leaving the disk due to the natal kick. A young pulsar with a significant height above the thin disk it likely moving in the same direction, i.e., away from the Galactic plane.  

Here, we consider a previously unappreciated selection effect (see, e.g., \citealt{Lorimer:1997} for a detailed discussion of other relevant selection effects in inferring pulsar velocities).    There is a growing body of observational evidence that spins and velocities of pulsars are preferentially aligned \citep{Johnston:2005,Wang:2006,Postnov:2008,Noutsos:2013,Yao:2021}.  Meanwhile, \citet{Janka:2022} proposed that supernova fallback would naturally explain the alignment of the spin axis with the kick direction, though the neutrino-induced kick models of \citet{ColemanBurrows:2022} show a weaker anti-alignment.  If pulsar spins are indeed aligned with their kicks, then the detectability of the pulsar, which depends on the location of the observer relative to the spin axis, implies that the radial component of the velocity distribution of {\it detectable} pulsars follows a different distribution than the components perpendicular to the line of sight.\footnote{Our assumption that the pulsar's velocity closely follows the natal kick is predicated on the fact that the typical kicks are much larger than the peculiar velocity of the progenitor star, including due to orbital motion if the pulsar is ejected from a wide binary \citep[e.g.,][]{Kapil:2022}, and that subsequent acceleration in the Galactic potential does not significantly change the velocity for young pulsars.} 

Here, we evaluate the implications of the anisotropy in the velocities of detectable pulsars on the inference of true pulsar speeds from their observed 2-d velocities.  Below, we employ toy models to demonstrate the impact of this anisotropy and estimate its potential effect.  We do not re-analyze the pulsar sample, which would require a choice of the shape of the kick velocity distribution (cf.~\citealt{Igoshev:2020,Kapil:2022}), a forward model for pulsar evolution in the Galaxy, and careful accounting for observational errors and selection effects \citep{Arzoumanian:2002}.

\section{Methodology and motivation}

\begin{figure}
\centering
\includegraphics[width=0.3\textwidth]{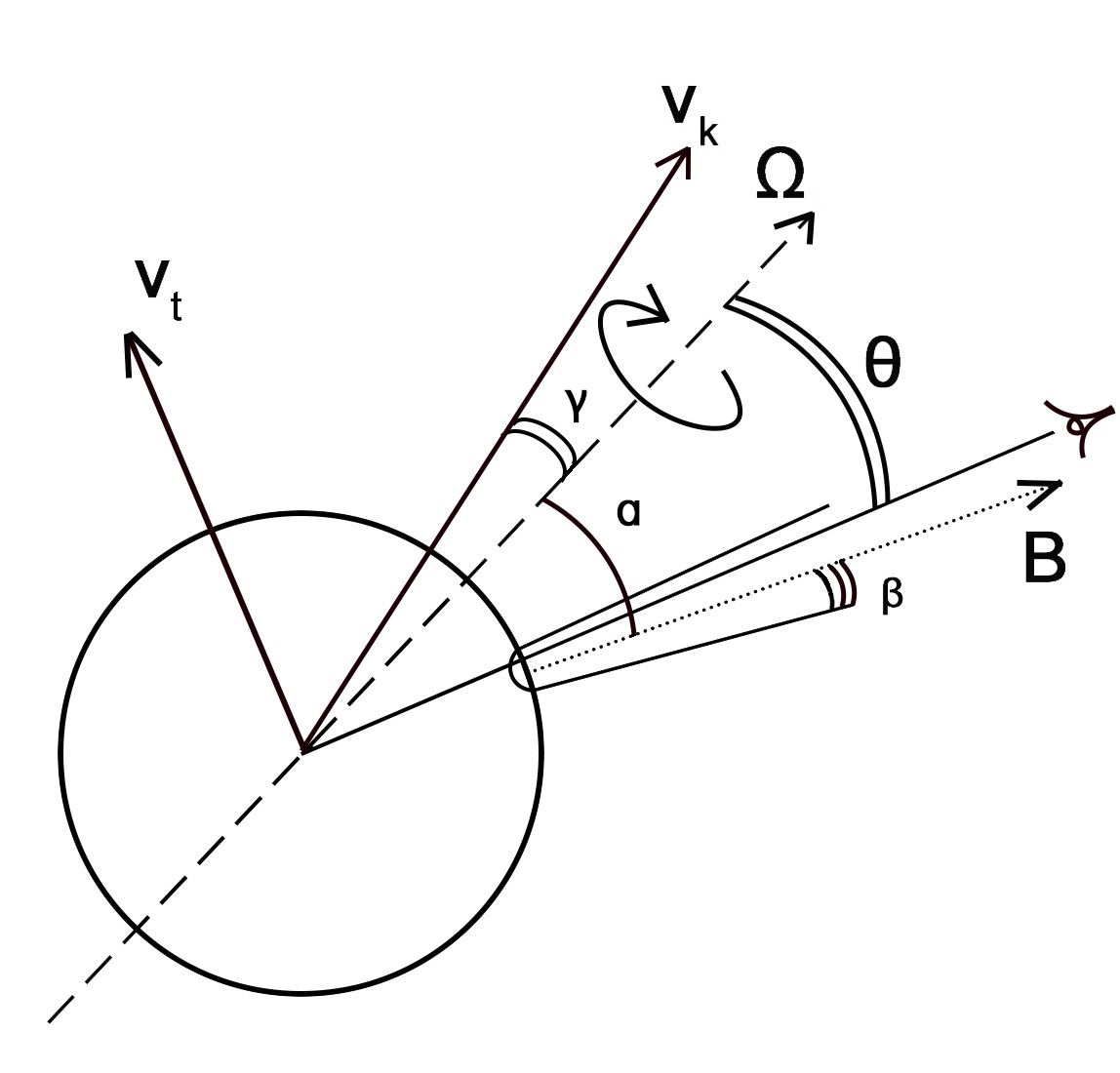}
\caption{\label{fig:sketch}Sketch of the angles in the problem.  The pulsar velocity $v_k$ is assumed to be misaligned from the spin axis by a small angle $\gamma$; 
$\alpha \leq \pi/2$ is the angle between the beam center and the spin axis; $\theta \leq \pi/2$ is the angle between the observer and the spin axism so that $v_t$ is the velocity transverse to the line of sight}; $\beta$ is the half-opening angle of each of the bipolar, conical beams.
\end{figure}

We assume that the pulsar velocity is approximately aligned with the spin vector, with misalignment  angle $\gamma \ll 1$ (this does not account for the acceleration of the pulsar in the Galactic potential and is therefore relevant only for young pulsars). We consider a simple bipolar conical beam, with half-opening angle $\beta \leq \pi/2$, whose axis is misaligned by an angle $\alpha \leq \pi/2$ from the spin axis.  We consider an observer viewing the pulsar from an angle $\theta$ off the spin axis (which can also be assumed to be $\leq \pi/2$ without loss of generality for a bipolar beam).  The setup is illustrated in Figure \ref{fig:sketch}.

In order for a radio source to be discovered and classified as a radio pulsar, it must have detectable {\it pulsed} emission.  We therefore require that the observer must at least sometimes, but not always, be inside the beam.  This is satisfied if and only if the following four conditions are met:
(i) $\theta < (\alpha+\beta)$, (ii) $\theta < \pi - (\alpha+\beta)$; (iii) $\theta > (\alpha-\beta)$; and (iv) $\theta > (\beta-\alpha)$.  (Note that several of these will be met trivially -- e.g., if $\alpha>\beta$, condition (iv) is automatically satisfied, while if $\alpha<\beta$, condition (iii) is automatically satisfied.)

We can now ask the question of how the 2-d observable pulsar velocity distribution would differ from the isotropically projected 2-d velocity distribution for different choices of $\alpha$ and $\beta$.  

Generally, we expect that if the beams are closely aligned with the spin axis ($\alpha \ll \pi/2$) and are moderately narrow (small $\beta$), only observers looking nearly along the spin axis (small $\theta$) will be able to detect the pulsar.  Consequently, if the spin is aligned with the velocity, such observers will see a significantly reduced projection of the total speed onto the plane of the sky.  The proper motions of detectable pulsars with closely aligned velocities and magnetic dipoles will nearly vanish.  The full speeds inferred from the 2-d speeds under the assumption of isotropy could then be greatly under-estimated.

On the other hand, if moderately narrow beams are significantly misaligned from the spin axis ($\alpha \to \pi/2$), only observers who are also at nearly right angles to the spin axis ($\theta \to \pi/2$) will be able to detect the pulsar.  Such observers would see nearly the full velocity vector projected onto the plane of the sky, so their observed 2-d velocity distribution will be greater than under the isotropic assumption.  The full pulsar speeds inferred from 2-d speeds under the assumption of isotropy would be over-estimated in this case. The over-estimate for $\alpha \to \pi/2$, $\beta \to 0$, $\gamma \to 0$ has a maximum of $\sqrt{3} / \sqrt{2} \approx 1.22$.

Moreover, narrower beams (smaller $\beta$) will show a stronger selection effect, which should be progressively washed out as the beam width increases (larger $\beta$).

\begin{figure}
\centering
\includegraphics[width=0.5\textwidth]{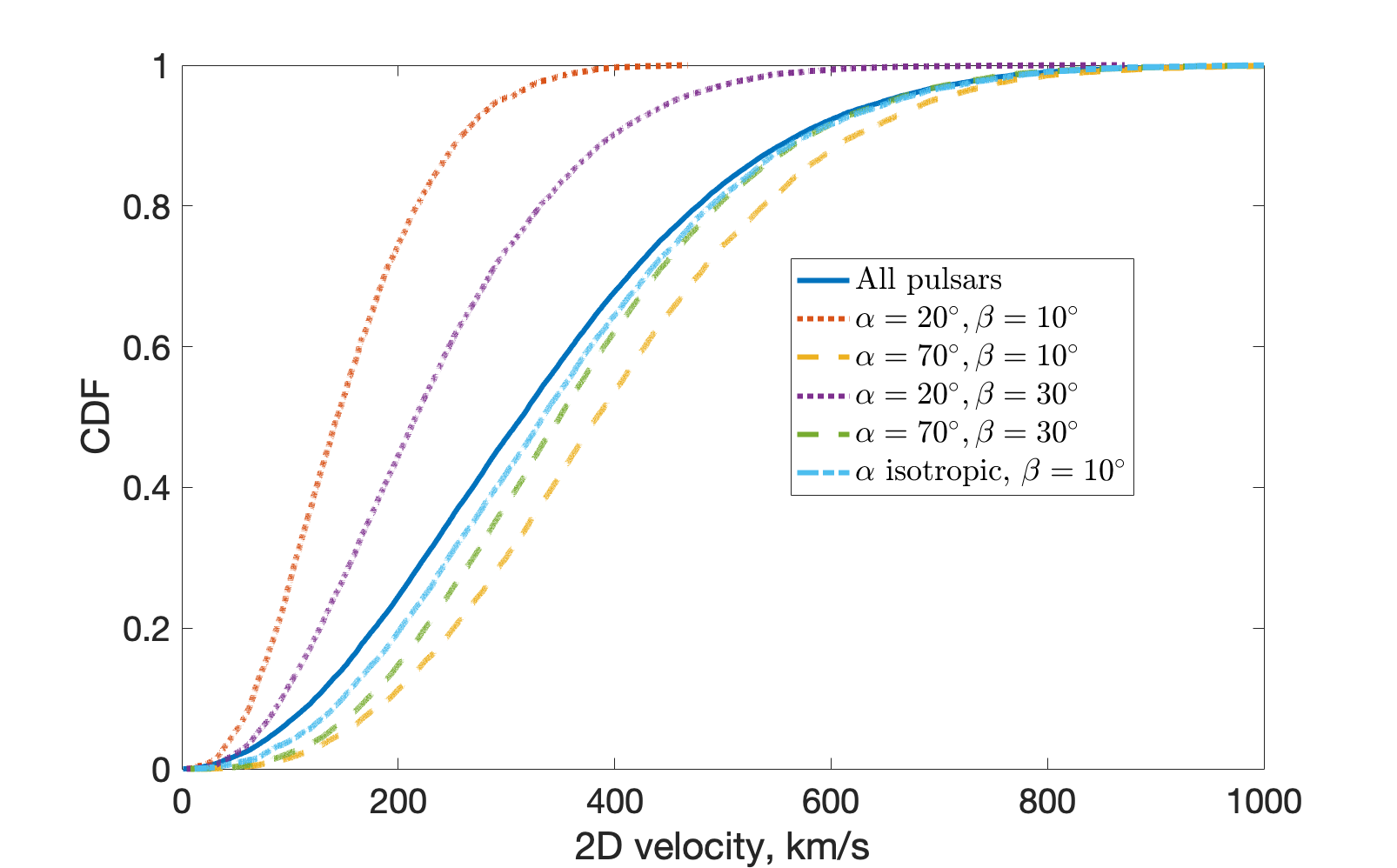}
\caption{\label{fig:velocityCDF} The cumulative distribution functions of 2-d velocity projections of all pulsars sampled from a Maxwellian distribution with $\sigma=265$ km s$^{-1}$ (dark blue) and the subset of detectable pulsars under varying assumptions about the beam misalignment angle $\alpha$ and the beam half-width $\beta$, with $\gamma=0$ (see text).}
\end{figure}

We illustrate these points in Figure \ref{fig:velocityCDF}.  For illustrative purposes only, we sample the 3-d speed follows a Maxwellian distribution with one-dimensional root-mean-squared velocity dispersion $\sigma=265$ km s$^{-1}$, following \citet{Hobbs:2005}.  We consider pulsar velocities perfectly aligned with the spin axis, $\gamma=0$.    The observer's location is not a priori special, so the angle $\theta$ is isotropically distributed, $p(\theta) = \sin \theta$.  We consider several different combinations of $\alpha$ and $\beta$.  We plot the cumulative distribution functions (CDFs) of projected 2-d velocities of all pulsars, ignoring selection effects, in blue.  Other CDF curves show the 2-d velocities for pulsars that are considered detectable according to the conditions specified above.

As expected, large misalignment angles $\alpha = 70^\circ$ (green and yellow dashed curves) shift the observed 2-d distribution upward, while small misalignment angles $\alpha = 20^\circ$ (red and purple dotted curves) shift it downward.  Moreover, in line with expectations the shifts are greater for narrow beams with $\beta=10^\circ$ (red, yellow) than wide beams with $\beta=30^\circ$ (purple, green).  In light blue, we show the case of narrow ($\beta=10^\circ$) beams with isotropically distributed misalignment angles, $p(\alpha) = \sin \alpha$.  Because this distribution prefers more misaligned beams, the 2-d velocities are shifted higher than under the assumption of no selection effects.  

\section{Simplified population synthesis}

In this section, we illustrate the impact of the anisotropy in the velocities of detectable pulsars with respect to the observer's line of sight on natal kick inference.  In order to do this, we assume that the true pulsar velocities follow a Maxwellian distribution.  This is partially motivated by, e.g., \citet{Hobbs:2005}, who fit all observed pulsar velocities to a Maxwellian with $\sigma=265$ km s$^{-1}$, and by \citet{Igoshev:2020}, who found that the high-velocity component of the pulsar distribution can be fit to a Maxwellian with $\sigma = 336$ km s$^{-1}$.  Conveniently, under the assumption of a Maxwellian distribution with isotropic velocities and the absence of measurement errors, the maximum-likelihood value of $\sigma$ can be inferred directly from the root-mean-square of the observed 2-d velocities as follows.

If measurements are perfect\footnote{Real pulsar velocity measurements are, of course, not perfect, nor are pulsar velocities drawn from a single Maxwellian distribution \citep[e.g.,][]{Igoshev:2020,Kapil:2022}, so this calculation is only intended to indicate the likely size of the effect.}, the likelihood of an individual observation is just the probability distribution from which it is drawn, assumed to be the two-dimensional Maxwellian probability distribution if the pulsar velocities are isotropic:
\begin{equation}
p(v_\mathrm{2D} | \sigma ) = \frac{v_\mathrm{2D}}{\sigma^2} \exp\left[ - \frac{v_\mathrm{2D}^2}{2\sigma^2} \right]\, .
\end{equation}
The likelihood of a set of independent two-dimensional pulsar velocity observations $\{v^i_\mathrm{2D}\}$ for $i \in [1,N]$ is given by the product of individual likelihoods, so the log-likelihood can be written as a sum:
\begin{equation}
\log \mathcal  L (\sigma) = \sum_{i=1}^N \log p(v^i_\mathrm{2D} | \sigma)\, .    
\end{equation}
In this simple case the log-likelihood can be optimised analytically,
\begin{equation}
\frac{\partial \log \mathcal L}{\partial \sigma} = -\frac{2N}{\sigma} + \frac{1}{\sigma^3} \sum_{i=1}^N (v^i_\mathrm{2D})^2 = 0\, .
\end{equation}
Hence the maximum likelihood estimate of $\sigma$ is
\begin{equation}
\hat{\sigma}= \sqrt{\frac{\langle v^2_\mathrm{2D} \rangle  }{2}}\, ,
\end{equation}
where $\langle v^2_\mathrm{2D} \rangle$ is the mean value of squared two-dimensional speeds. 

In order to evaluate the impact of anisotropy among detectable pulsars, we draw a mock population of pulsar with natal kicks sampled from a Maxwellian distribution with parameter $\sigma$, determine if they are detectable as described in the previous section from the perspective of an observer placed at a random isotropically chosen observer angle $\theta$, $p(\theta) = \sin \theta$, compute the transverse velocities $v_\mathrm{2D} = v \sin \theta$ for the detectable pulsars, and evaluate the systematic bias in the inferred maximum-likelihood parameter $\hat{\sigma}$ under the assumption of isotropy in the pulsar velocity, $\hat{\sigma}/\sigma$.  

\begin{figure}
\centering
\includegraphics[width=0.5\textwidth]{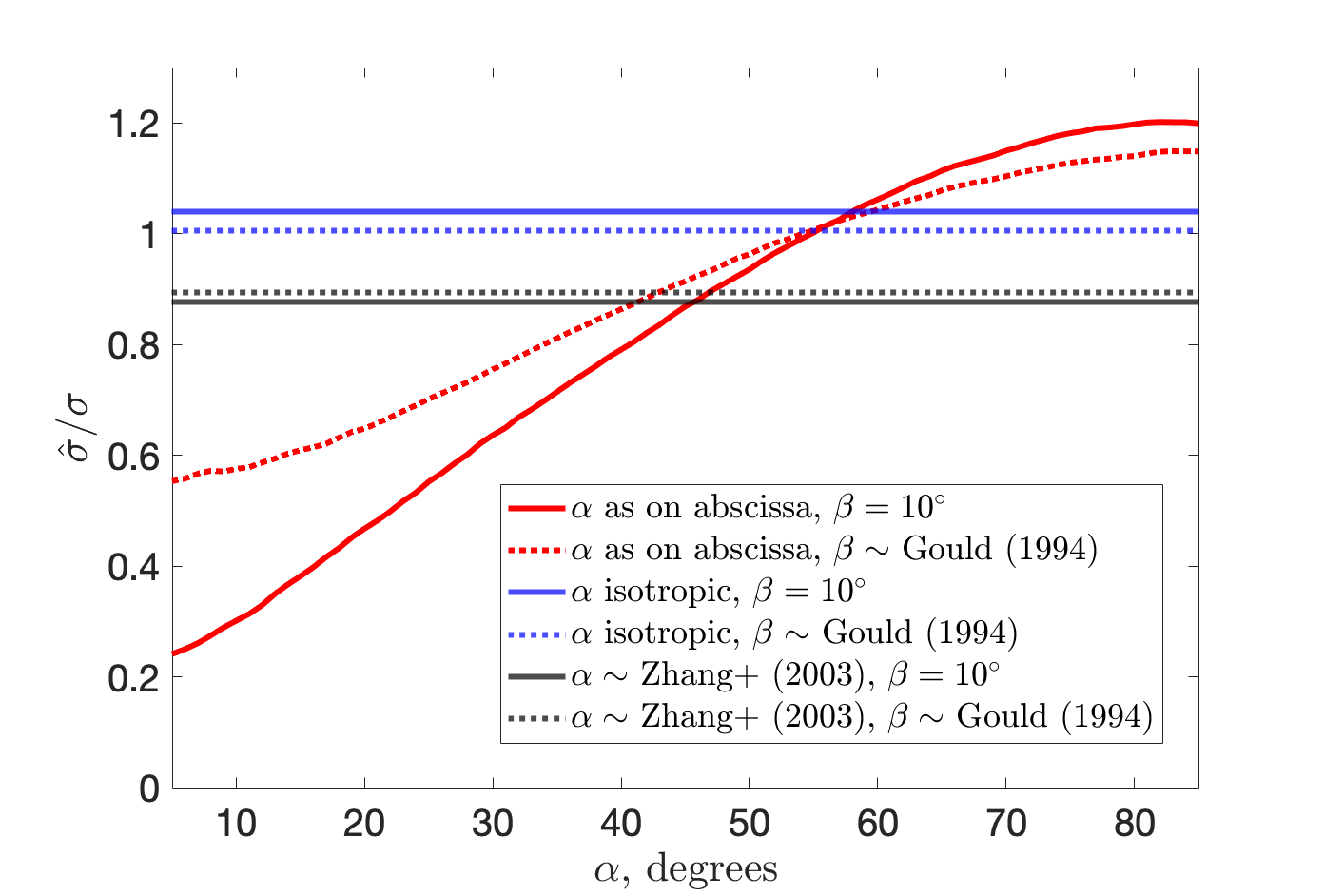}
\caption{\label{fig:inference} The multiplicative bias of the inferred Maxwellian velocity distribution parameter $\hat{\sigma}$ relative to its true value $\sigma$ when an isotropic projection of 3-d velocity is assumed for inference, without accounting for velocity-direction-correlated selection effects.}
\end{figure}

In Figure \ref{fig:inference}, we plot the bias $\hat{\sigma}/\sigma$ as a function of the misalignment angle $\alpha$ for $\beta=10^\circ$ in solid red.  As suggested by Fig.~\ref{fig:velocityCDF} and the associated discussion, $\alpha$ has a very strong impact on the 2-d velocities of detectable pulsars, with larger $\alpha$ leading to larger projected velocities and, hence, over-estimation of the true velocity dispersion when an isotropic velocity distribution is assumed.  

We indeed find that large misalignments between the beam and the spin/kick axis ($\alpha \to \pi/2$) lead to a $\sim 15\%$ systematic over-estimate of the underlying kick velocity as parametrised by $\sigma$.  Conversely, low misalignment angles $\alpha \to 0$ could lead to the underlying natal kick velocity distribution being under-estimated by up to a factor of 4, as most detectable pulsars would be viewed nearly on-axis, with transverse velocities strongly suppressed.

Of course, in a realistic pulsar population $\alpha$ and $\beta$ are not fixed, but follow some distributions, which we assume are independent of each other and of the kick velocity \citep[but see, e.g.,][for further discussions]{FGKaspi:2006}.  For the beam radius $\beta$, we use a relationship between $\beta$ and the initial pulsar period $P_0$ fit by \citet{Gould1994PhDT} (see also \citealt{TaurisManchester1998MNRAS}):
\begin{equation}
\beta (P_0) = 5^\circ.4 \left( \frac{P_0}{1\; \mathrm{sec}} \right)^{-1/2}\, .
\label{e:beta_p0}
\end{equation}
We follow \citet{Igoshev2022MNRAS} in drawing the periods of young pulsars from a log-normal distribution with $\mu_{\log (P/\mathrm{s})} = -1.04$ and $\sigma_{\log P} = 0.53$ for periods in the range $P_0\in[0.01, 2]$~s.  This leads to a mean $\langle\beta\rangle \approx 20^\circ$, larger than the fixed value of $10^\circ$ assumed previously.  Following the analysis in the previous section, we expect larger $\beta$ to reduce the bias in the inferred velocity, which is exactly what we see in Fig.~\ref{fig:inference}, where dotted curves indicate $\beta$ sampled according to Eq.~(\ref{e:beta_p0}).

We also consider two distributions for the obliquity angle $\alpha$.    The systematic bias for a population with an isotropic distribution of $\alpha$ is indicated with a blue horizontal line in Fig.~\ref{fig:inference}.  As expected from the minimal shift in the 2-d velocity CDFs in Fig.~\ref{fig:velocityCDF}, an isotropic $\alpha$ distribution leads to very small biases  of 4\% (0.5\%) in the recovered $\sigma$ for $\beta=10^\circ$ ($\beta$ from Eq.~\ref{e:beta_p0}).

The distribution of the obliquity angle $\alpha$ has been estimated for hundreds of radio pulsars \citep{LyneManchester988MNRAS,Rankin1993ApJ,Rankin1993ApJS}.  \citet{TaurisManchester1998MNRAS} combined these measurements to argue that the distribution of $\alpha$ is not uniform and peaks at lower values.  \citet{ZhangJiang2003PASJ} proposed the following probability density function:
\begin{equation}
p(\alpha) = \frac{0.784}{\cosh \left(3.5 [\alpha - 0.43]\right)} + \frac{0.294}{\cosh\left(4[\alpha - 1.6]\right)}\, ,   
\label{e:alpha}
\end{equation}
where $\alpha \in [0, \pi / 2]$.  This fit does indeed support lower values of $\alpha$, with a peak around 25$^\circ$, and we may thus expect that a population that follows this $\alpha$ distribution would lead to an underestimate of the pulsar velocity distribution.  Indeed, we find that the velocity distribution is underestimated by 12\% (11\%) for $\beta=10^\circ$ ($\beta$ from Eq.~\ref{e:beta_p0}) as shown in black in Fig.~\ref{fig:inference}.

Yuri Levin (priv.~comm.) suggested a possible test of the consistency of this model.  Pulsars with lower 2-d velocities should be the ones preferentially viewed close to the spin axis in this model.  They should then have larger pulse duty cycles.  However, we find that our theoretical models predict only moderate anti-correlations between the 2-d observed velocities and the pulse duty cycle, estimated as $\sqrt{\beta^2-(\alpha-\theta)^2}/(\pi \sin\theta)$.  For example, we predict the Pearson correlation coefficient between the pulse duty cycle and 2-d speed of detectable pulsars to be $-0.3$ for the \citet{ZhangJiang2003PASJ} $\alpha$ distribution and the $\beta$ distribution from Eq.~\ref{e:beta_p0}, but nearly zero for $\alpha=75^\circ$ and $\beta=10^\circ$.  Moreover, the sampling scatter in this correlation coefficient from a small sample of 80 pulsars -- the number for which VLBI parallax and proper motions data are available (see table A1 in \citealt{Igoshev:2020}) -- is quite large (e.g., 0.13 in the latter example).  The model is thus consistent with the measured correlation coefficient of $-0.13$ between the pulse duty cycle and the 2-d velocities of pulsars with VLBI data.  Here, the pulse duty cycle is computed as the ratio between the width at 50\% amplitude, obtained from the ATNF pulsar catalogue\footnote{\url{http://www.atnf.csiro.au/research/pulsar/psrcat}, v.~1.68} \citep{atnf}, and the pulsar period.  While consistent, this test is inconclusive.

\section{Implications}

We have shown that if young pulsars have natal kicks aligned (or anti-aligned) with their spins, then the distribution of observed 2-d velocities of detectable pulsars may be systematically shifted relative to isotropic projections of pulsar 3-d speeds onto the plane of the sky (see Figure \ref{fig:velocityCDF}).  Consequently, the 3-d speed distribution inferred from the observed 2-d velocities could be systematically biased (see Figure \ref{fig:inference}).  

The size of the bias strongly depends on the distribution of the pulsar obliquity angle $\alpha$ and pulsar beam size $\beta$.   A preference for small obliquity angles found by \citet[e.g.,][]{TaurisManchester1998MNRAS} would suggest that pulsar 3-d speeds are $\sim 10\%$ larger than inferred in previous studies, based on the \citet{ZhangJiang2003PASJ} fit to the $\alpha$ distribution.  However, this fit was incorrectly derived by using the Kolmogorov-Smirnov test statistic as a form of likelihood.  Of course, this test should only be used to accept or reject a particular hypothesis and cannot be used to compare acceptable hypotheses against each other.  

Perhaps a more serious issue is that both $\alpha$ and $\beta$ appear to evolve as the pulsar ages.  For example, figures 5 and 6 of \citet{TaurisManchester1998MNRAS} show, based on the data of \citet{Gould1994PhDT,Rankin1993ApJ}, that pulsars become more aligned over time, matching theoretical expectations \citep[e.g.,][]{Goldreich:1970,Philippov:2014}.  Conversely, young pulsars should have larger misalignment angles than the overall pulsar population.  Studies of pulsar kick angles typically rely on young pulsars \citep[e.g.,][]{Igoshev:2020,Willcox:2021,Kapil:2022}, which may thus have preferentially larger $\alpha$, possibly leading to a $\sim 15\%$ over-estimate in the inferred kick velocities after accounting for the wider beams of younger pulsars  \citep{Gould1994PhDT,TaurisManchester1998MNRAS}.  This effect could be reduced if the alignment between pulsar kicks and spins is imperfect ($\gamma > 0$).

Velocities are inferred from the analysis of young pulsars because older pulsars exhibit signs of acceleration in the Galactic potential, so their current velocities no longer reflect their natal velocity kicks.  This also means that the effect we describe in this paper would disappear for older pulsars, as pulsar acceleration would change the velocity orientation relative to the original kick direction while leaving the spin direction intact, destroying the spin-kick alignment.  Indeed, \citet{Noutsos:2013} noted that the correlation between the orientation of the proper motion and spin axis of a pulsar disappears for objects with kinematic ages above $10$~Myr. 

Furthermore, these considerations apply only to non-recycled pulsars.  The spin direction of a recycled pulsar is set by the orbital angular momentum of its host binary.  While the velocities of such binaries, or pulsars ejected from the binaries, may be correlated to the orbital plane (e.g., some kick directions are more or less likely to disrupt the binary), this is likely a less significant effect.

In summary, if young pulsars have kicks preferentially aligned with their spins, large obliquities and relatively narrow beams, then their 3-d speeds may typically be $\sim 15\%$ lower than the values inferred from observed 2-d transverse velocities without considering the effect discussed in this paper.  Reduced natal kicks would make it easier to retain the observed pulsar population in globular clusters and preserve a higher fraction of neutron-star binaries, including X-ray binaries and future gravitational-wave sources, from disruption during supernovae.

\acknowledgements
We thank Yuri Levin and Matthew Bailes for useful discussions. I.M.~acknowledges support from the Australian Research Council Centre of Excellence for Gravitational  Wave  Discovery  (OzGrav), through project number CE17010004. I.M.~is a recipient of the Australian Research Council Future Fellowship FT190100574. Part of this work was performed at the Aspen Center for Physics, which is supported by National Science Foundation grant PHY-1607611, with I.M.'s participation partially supported by the Simons Foundation. The work of A.P.I.~is supported by STFC grant no.\ ST/S000275/1.

\bibliographystyle{hapj}
\bibliography{Mandel,Igoshev}

\end{document}